\documentclass[aps,prb,showpacs,twocolumn,reprint,superscriptaddress]{revtex4-1}
\usepackage{amsmath}
\usepackage{amssymb}
\usepackage{graphicx}
\usepackage{color}
\usepackage{multirow}
\usepackage{dcolumn}
\usepackage{textcomp}
\begin{document}
\title{Effects of magnetic dopants in (Li$_{0.8}$M$_{0.2}$OH)FeSe (M = Fe, Mn, Co): a density-functional theory study using band unfolding technique}
\author {M. X. Chen}
\affiliation{College of Physics and Information Science, Hunan Normal University, Changsha, Hunan 410018, China}
\affiliation{Department of Physics, University of Wisconsin, Milwaukee, Wisconsin 53211, USA}
\author {Wei Chen}
\affiliation{Department of Physics and School of Engineering and Applied Sciences, 
Harvard University, Cambridge, Massachusetts 02138, USA}

\affiliation{International Center for Quantum Design of Functional Materials (ICQD), 
Hefei National Laboratory for Physical Sciences at Microscale, and 
Synergetic Innovation Center of Quantum Information and Quantum Physics, 
University of Science and Technology of China, Hefei, Anhui 230026, China}
\author {Zhenyu Zhang}
\affiliation{International Center for Quantum Design of Functional Materials (ICQD), 
Hefei National Laboratory for Physical Sciences at Microscale, and 
Synergetic Innovation Center of Quantum Information and Quantum Physics, 
University of Science and Technology of China, Hefei, Anhui 230026, China}
\author {M. Weinert}
\affiliation{Department of Physics, University of Wisconsin, Milwaukee, Wisconsin 53211, USA}

\date{\today}

\begin{abstract}
The effects of Fe dopants on the electronic bands structure of (Li$_{0.8}$Fe$_{0.2}$OH)FeSe
are investigated by a band unfolding ($k$-projection)
technique and first-principles supercell calculations.
Doping 20\% Fe into the LiOH layers causes electron donation to the FeSe layers, significantly 
changing the profile of bands around the Fermi level.
Because of the weak bonding between the LiOH and FeSe layers 
the magnetic configuration of the dopants has only minor effects on the band structure.
The electronic bands for the surface FeSe layer of (Li$_{0.8}$Fe$_{0.2}$OH)FeSe show noticeable differences compared
to those of the inner layers, both 
in the location of the Fermi level and in details of the bands near the high symmetry points, resulting from different effective
doping levels and the broken symmetry at the surface.
The band structure for the surface FeSe layer with checkerboard antiferromagnetic order 
is reasonably consistent with angle-resolved photoemission results.
The 3$d$ transition-metals Mn and Co have similar doping effects on the band structure of (LiOH)FeSe.

\end{abstract}

\pacs{71.20.-b,73.20.-r,73.22.Pr}

\keywords{FeSe; band unfolding}

\maketitle
\section{INTRODUCTION}
The discovery of superconductivity in REFeAsO (RE = rare earth) compounds with T$_\mathrm{c}$ as high as
55 K has engendered significant interest in searching for new Fe-based superconductors.
One surprise from these recent explorations is that an FeSe monolayer epitaxially grown on SrTiO$_3$
shows superconducting T$_\mathrm{c}$ as high as 77 K,\cite{qing-yan_2012} much higher than its bulk phase.
One unusual and puzzling feature of this system is that its Fermi surface is characterized 
by an electron-like pocket centered around the M points, while no pocket appears near the $\Gamma$ point.
\cite{liu_electronic_2012,he_phase_2013,tan_2013,liu_dichotomy_2014} 
This observation challenges the Fermi surface nesting scenario that relies on nesting between M and $\Gamma$-point electronic states. 
\cite{mazin_2008,kuroki_2008}
More recently, a superconducting T$_\mathrm{c}$ over 40 K was observed in an FeSe-based bulk phase, (Li$_{1-x}$Fe$_{x}$OH)FeSe.
\cite{lu_coexistence_2015,pachmayr_coexistence_2014,dong_phase_2015,dong_2015_PRB}
Unlike the binary FeSe compound, (Li$_{1-x}$Fe$_{x}$)OH layers are intercalated between the FeSe layers.
This new type of FeSe-based superconductor is also distinct from the alkali-metal intercalated counterpart: 
it is stable at ambient condition, whereas the later is extremely sensitive to air.
Moreover, magnetism was found to coexist with superconductivity in Li$_{1-x}$Fe$_{x}$OH.
\cite{lu_coexistence_2015,pachmayr_coexistence_2014} 
The angle-resolved photoemission spectroscopy (ARPES) experiments of Ref.~[\onlinecite{Feng_PRB_2015}] find that 
(Li$_{0.8}$Fe$_{0.2}$OH)FeSe has an electronic structure akin to that of Rb$_{0.76}$Fe$_{1.87}$Se$_{2}$,
with similar band dispersions and gap symmetry.
In addition, ARPES experiments find that
the band structure for (Li$_{0.8}$Fe$_{0.2}$OH)FeSe shares great similarities to that of FeSe/SrTiO$_3$, 
suggesting that the superconductivity in this system and in FeSe/SrTiO$_3$ may have a common electronic origin.\cite{zhao_common_2016}

Several density functional theory (DFT) calculations have been performed for (Li$_{1-x}$Fe$_x$OH)FeSe.
\cite{zhao_common_2016,Nekrasov2015,Chenwei_PRB_2015}
However, the reported DFT-derived band structures are for nonmagnetic (LiOH)FeSe without the Fe dopants, 
and are found to be largely inconsistent with the ARPES measurements.\cite{zhao_common_2016,Nekrasov2015} 
In particular, hole pockets near the $\Gamma$ point seen in the DFT calculations, 
are actually absent in the ARPES measurements.\cite{zhao_common_2016,Nekrasov2015}
DFT supercell calculations carried out by some of us\cite{Chenwei_PRB_2015} find that
the Fe dopants play an important role in the structural stability of (Li$_{0.8}$Fe$_{0.2}$OH)FeSe, and 
the charge transfer between the LiOH and FeSe layers.
Although both Refs.~[\onlinecite{lu_coexistence_2015}] and [\onlinecite{pachmayr_coexistence_2014}]
find that the magnetism coexisting with the superconductivity originates from the Fe dopants,
they report different kinds of magnetism, i.e., ferromagnetism vs antiferromagnetism.
This discrepancy was also seen in the DFT calculations.
The local-density approximation (LDA) calculations of
Ref.~[\onlinecite{Nekrasov2015}] give ferromagnetically aligned Fe moments in the LiOH layers, while
 our calculations\cite{Chenwei_PRB_2015} using the Perdew-Burke-Ernzerhof GGA find 
that they are antiferromagnetically aligned.
The different magnetic orderings found may originate from different configurations/distributions of Fe dopants since
the energy difference between the AFM and FM coupling of the Fe dopants is small.\cite{Chenwei_PRB_2015}
Moreover, our study\cite{Chenwei_PRB_2015} predicted that (Li$_{0.8}$Co$_{0.2}$OH)FeSe is structurally stable with even
larger electron injection, possibly leading to a higher T$_\mathrm{c}$.
Intrinsically, spin-orbit coupling (SOC) also plays an important in the electronic properties of Fe-based superconductors,
\cite{hao_2014,soc-2015,borisenko_2016}
including driving the FeSe monolayers into nontrivial topological phases.\cite{hao_2014}
Dopants such as oxygen vacancies in the SrTiO$_3$ substrate also can affect the SOC spin splitting at the M point for the
checkerboard antiferromagnetic state.\cite{chen_2016_PRB}
As a step to gaining a better understanding of the physics underlying the superconductivity of these systems,
it is worthwhile investigating the effects of the dopants, magnetic ordering, SOC, and the interplay among them on
the normal state band structure
of (Li$_{0.8}$Fe$_{0.2}$OH)FeSe.

In this paper, we investigate the band structure of (Li$_{0.8}$Fe$_{0.2}$OH)FeSe by carrying out DFT supercell calculations.
The effects of the Fe dopants on the electronic bands are studied by means of a band unfolding technique that 
projects the supercell wave functions onto the $k$-points in the Brillouin zone of the 1$\times$1 chemical unit cell of (LiOH)FeSe.
We find that doping Fe into the LiOH layers not only shifts the Fermi level, 
but also induces significant changes in the profile of bands around the Fermi level.
The magnetic ordering of the dopants in the LiOH layers, however, has only minor effects on the band structure.
The band structure for the surface FeSe layer shows significant differences from those for inner layers in both 
the location of the Fermi level and band details near high symmetry points.
We further investigate the band structures for Mn- and Co-doped (Li$_{0.8}$M$_{0.2}$OH)FeSe (M = Mn and Co),
and show that Fe, Mn, or Co doping has similar effects on the band structure of (LiOH)FeSe. 

\section{COMPUTATIONAL DETAILS}
\begin{figure}
  \includegraphics[width=0.48\textwidth]{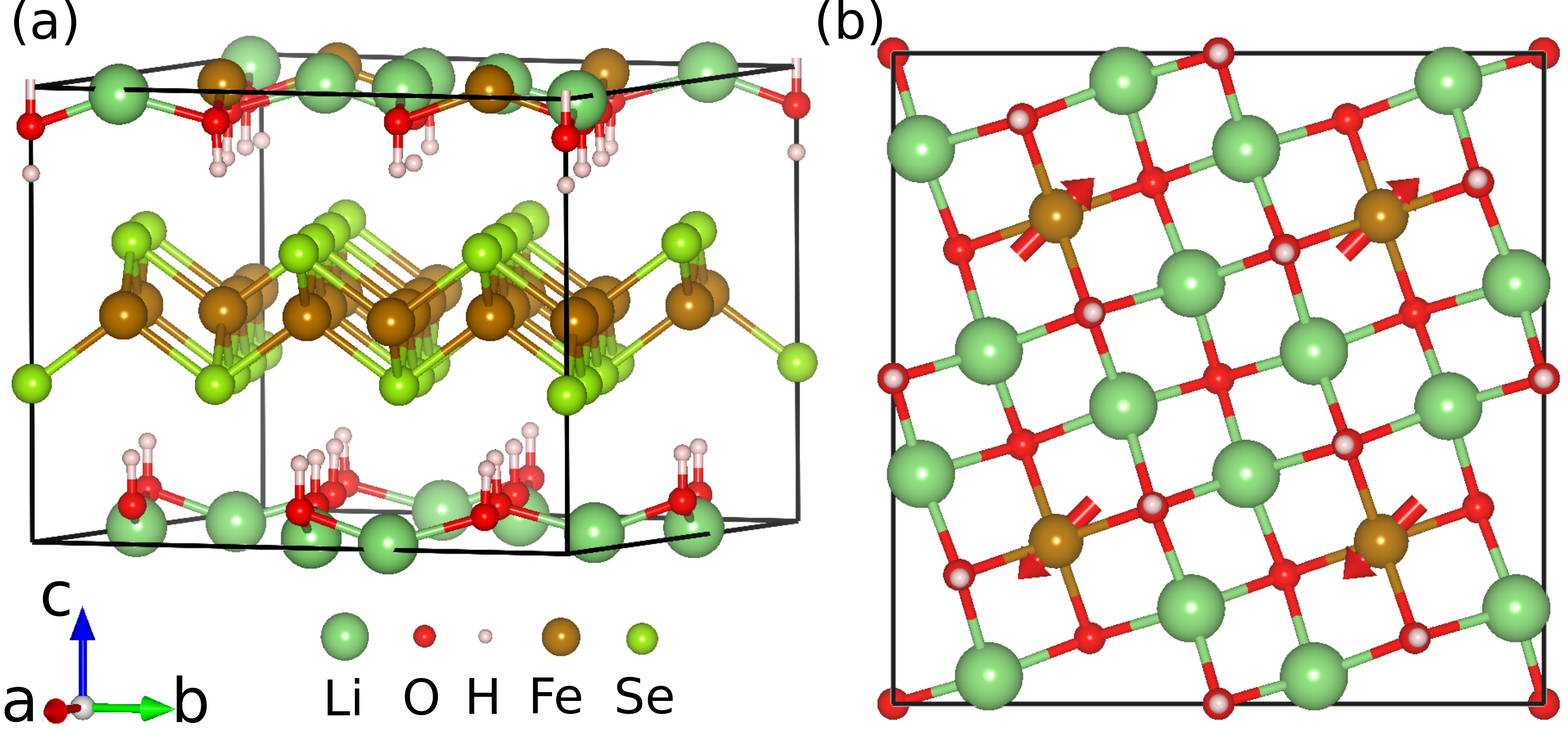}
  \caption{Structural model of (Li$_{0.8}$Fe$_{0.2}$OH)FeSe.
  (a) Perspective view of (Li$_{0.8}$Fe$_{0.2}$OH)FeSe. 
  (b) Top view of Li$_{0.8}$Fe$_{0.2}$OH layer. 
  The red arrows represent the relative magnetization of the Fe dopants in the collinear antiferromagnetic configuration.
  In addition to the collinear AFM configuration, 
  checkerboard antiferromagnetic and ferromagnetic between the Fe dopants (not shown) are also investigated in the present study.
  }
 \label{fig1}
\end{figure}
Our calculations were performed using the Vienna Ab Initio Simulation Package. \cite{kresse_efficiency_1996,kresse_efficient_1996}
Our previous study examined the effects of different exchange-correlation functionals on the lattice parameter
and found that the Perdew-Burke-Ernzerhof (PBE) functional without 
including the popular vdW corrections yields a reasonable description of the structural properties.\cite{Chenwei_PRB_2015}
Therefore, in the present study the exchange-correlation functional is approximated by the generalized gradient approximation as
parametrized by PBE,\cite{perdew_1996} and the pseudopotentials were
constructed by the projector augmented wave method.\cite{bloechl_projector_1994,kresse_ultrasoft_1999} 
A 3$\times$3$\times$1 Monkhorst-Pack $k$-mesh was used to sample the BZ and 
a plane-wave energy cutoff of 400 eV was used for electronic structure calculations. 
For calculations of (Li$_{0.8}$Fe$_{0.2}$OH)FeSe(001) a slab consisting of three FeSe layers and two
Li$_{0.8}$Fe$_{0.2}$OH layers was used, with the surface terminated by FeSe layers. 
The slab is separated from its periodic images by 10 \AA{} vacuum regions and was fully relaxed 
with a threshold of 0.001 eV/\AA{} for the residual force on each atom.  

We adapt the structural models reported in Ref.~[\onlinecite{Chenwei_PRB_2015}]
for (Li$_{0.8}$Fe$_{0.2}$OH)FeSe (Fig.~\ref{fig1}),
a $\sqrt{10}\times\sqrt{10}$ supercell of (LiOH)FeSe
with four Li atoms replaced by Fe atoms. 
To eliminate the band foldings caused by the use of supercells,
we project the supercell wave functions onto the corresponding momentum vector $k$ of the primitive (chemical) unit cell
\cite{bufferlayer,chen_revealing_2014}
of (LiOH)FeSe (two Fe atoms in the FeSe layer). 
In this way, the Fe dopants serve as a perturbation on the bands of (LiOH)FeSe,
whose effect can be seen in $k$-projected weights
given by $|\psi_\mathbf{k}(\mathbf{r})|^2$, 
where the wave function $\psi_\mathbf{k}$ is $k$-projected.
In essence, the procedure is to find
which $\mathbf{k}_p$ of the primitive cell each plane wave $e^{i({\mathbf{k}_s+\mathbf{G}})\cdot{\mathbf{r}}}$
belongs to, i.e., for integers $M_i$ and $m_j$, determining the fractional part $\kappa_j$ that
defines $\mathbf{k}_p$ of the primitive cell relative to $\mathbf{k}_s$
of the supercell: 
\begin{eqnarray*}
 \mathbf{G} &=& \sum_i M_i \mathbf{B}_i  = \sum_j (m_j + \kappa_j ) \mathbf{b}_j \\
            &=& \sum_j \left( \sum_i M_i (
\mathbf{B}_i\cdot\mathbf{a}_j ) \right)  \mathbf{b}_j ,
\end{eqnarray*}
with $\mathbf{a}_i\cdot\mathbf{b}_j=\delta_{ij}$,
where $\mathbf{B}$ and $\mathbf{b}$ are the reciprocal lattice vectors of the supercell and the primitive cell, respectively.

\section{RESULTS AND DISCUSSIONS}
\subsection{Unfolded bands of (Li$_{0.8}$Fe$_{0.2}$OH)FeSe}
\begin{figure*}
  \includegraphics[width=0.99\textwidth]{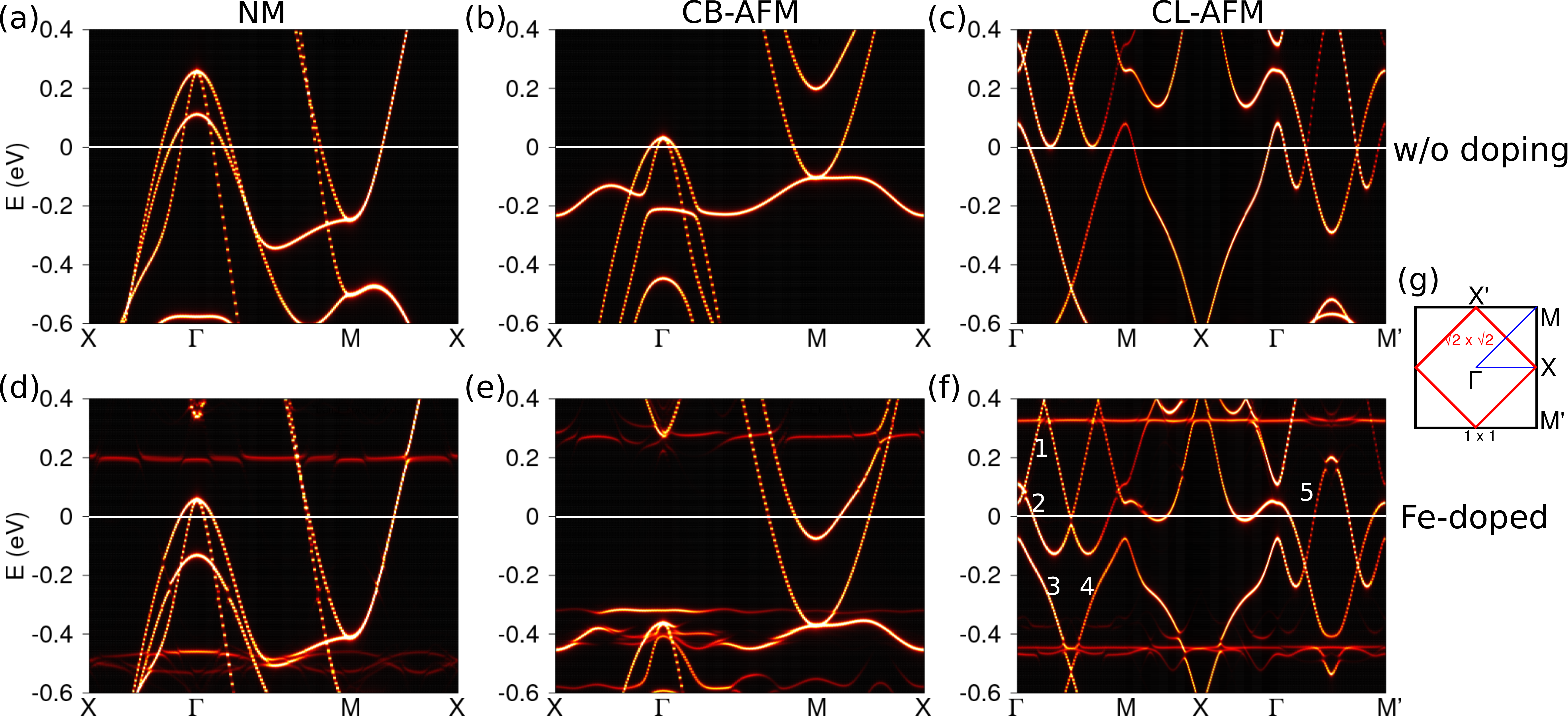}
  \caption{$k$-projected bands of (a)-(c) (LiOH)FeSe and (d)-(f) (Li$_{0.8}$Fe$_{0.2}$OH)FeSe along high-symmetry lines of the BZ of the 1 $\times$ 1 unit cell shown in (g). 
  The bands for different assumed magnetic configurations of the FeSe layer -- non-magnetic (NM), checkerboard antiferromagnetic (CB-AFM), and collinear
antiferromagnetic (CL-AFM) -- are shown; the Fe dopants have the CL-AFM order (Fig.~\ref{fig1}(b)).
  SOC was not included in these calculations and the Fermi level is set to zero.
  The numbers in (f) denote band indexes.
  (g) The BZs of the 1$\times$1 unit cell (black) and the $\sqrt{2}\times\sqrt{2}$ supercell (red).
  }
 \label{fig2}
\end{figure*}
We considered three magnetic configurations for the FeSe layers:
 non-magnetic (NM), checkerboard antiferromagnetic (CB-AFM), and collinear antiferromagnetic (CL-AFM).
For the Fe dopants in Li$_{0.8}$Fe$_{0.2}$OH, we considered the CB-AFM, CL-AFM, and ferromagnetic (FM) orderings.
The magnetic ground state is found to have CL-AFM ordering of the Fe atoms in both the FeSe and Li$_{0.8}$Fe$_{0.2}$OH layers,
with the interlayer spins aligned parallel to each other.\cite{Chenwei_PRB_2015}

Figure \ref{fig2} shows the $k$-projected bands for (LiOH)FeSe with and without the Fe dopants, 
for CL-AFM ordered Fe dopants.
The effects of the magnetic configuration of the Fe dopants on the band structure of (Li$_{0.8}$Fe$_{0.2}$OH)FeSe will be discussed later.
One prominent feature is that the Fermi level experiences an upward shift upon Fe doping.
For the NM state the shift is about 0.2 eV.
As a result, the hole pockets at $\Gamma$ shrink dramatically but remain finite at E$_\mathrm{F}$,
while the electron pockets at M enlarge significantly. 
For the CB-AFM ordering, the doping leads to the disappearance of the hole pockets at $\Gamma$.
The electron pocket at M enlarges and another electron pocket appears.
A comparison of Figs.~\ref{fig2}(b) and (e) shows that the hole-like bands at $\Gamma$ are shifted down to lower energy by about 0.1 eV
relative to the bottom of the electron band at M, reflecting the electronic response of the FeSe layer
to the electron doping into the system.
Consequently, the bandwidth is significantly reduced, 
from about 0.27 eV for the undoped (LiOH)FeSe to about 0.1 eV for (Li$_{0.8}$Fe$_{0.2}$OH)FeSe.  
This band renormalization is similar to that seen in FeSe/SrTiO$_3$(001) with interfacial oxygen vacancies.\cite{chen_2016_PRB}
(The flat bands, at around $\pm 0.3$ eV for CB-AFM, are due to the Fe dopants.)

The Fe dopants also affect the band structure for the CL-AFM state.
While the chemical and magnetic unit cells for the FeSe CB-AFM ordering are the same, for the CL-AFM configuration the
magnetic cell is a doubled $\sqrt{2}\times\sqrt{2}$ supercell. Because the magnetic effects on the bands are strong, the
bands $k$-projected onto the $1\times 1$ cell reflect the magnetic symmetry, including the repetition/folding of the bands
along $\Gamma$-M (M'). (Both M and M' of the $1\times 1$ BZ fold back to $\Gamma$ of the $\sqrt{2}\times\sqrt{2}$ BZ;
the $\Gamma$-M and $\Gamma$-M' directions differ because they correspond to $k$ parallel or perpendicular to the ferromagnetic
stripes of Fe atoms.) The intrinsic $1\times 1$ chemical symmetry is reflected in the intensity differences even for the undoped
system.
For example, there is an intensity difference between bands 3 and 4 along $\Gamma$-M for both (LiOH)FeSe and
(Li$_{0.8}$Fe$_{0.2}$OH)FeSe, and similarly for bands 1 and 2, as well as band 5, 
where the band intensity changes around the mid-point of $\Gamma$-M'.

A comparison of Figs.~\ref{fig2}(c) and (f) shows that
doping 0.2 Fe to (LiOH)FeSe shifts E$_\mathrm{F}$ by about 0.2 eV when the top of band 3 at $\Gamma$ is taken as the reference.
Similar to the CB-AFM case, the electron doping of the FeSe layer due to the Fe dopants induces strong band renormalization.
For example, (LiOH)FeSe band 2 has a width of about 0.35 eV from its maximum at $\Gamma$ to the local minimum at the $k$-point
between  $\Gamma$ and M, while it is $\sim$0.2 eV for (Li$_{0.8}$Fe$_{0.2}$OH)FeSe.
Similar band renormalizations can be observed for other bands as labeled in Fig.~\ref{fig2}(f).
These results suggest that the effects of the Fe dopants in the LiOH layers are similar to 
those of interfacial oxygen vacancies in FeSe/SrTiO$_3$.\cite{chen_2016_PRB}

\begin{figure}
  \includegraphics[width=0.30\textwidth]{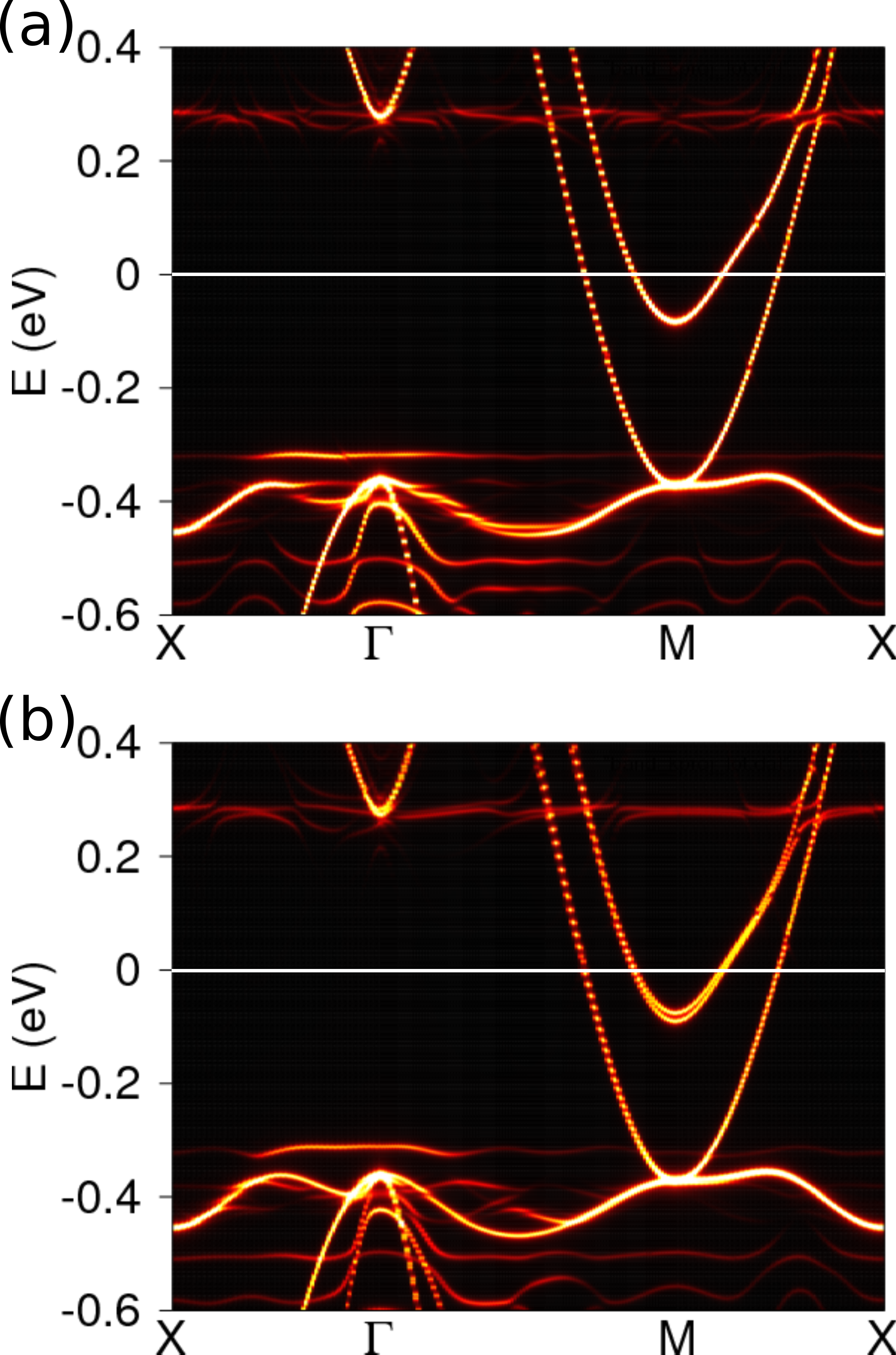}
  \caption{$k$-projected bands of (Li$_{0.8}$Fe$_{0.2}$OH)FeSe for different magnetic configurations of the Fe dopants:
  (a) CB-AFM and (b) FM. 
  The spins in the FeSe layers have the CB-AFM ordering.  
  }
 \label{fig3}
\end{figure}

To see how the magnetic configuration of the Fe dopants affects the electronic bands of (Li$_{0.8}$Fe$_{0.2}$OH)FeSe,
calculations were also performed for CB-AFM and FM ordered dopants. The bands, with the FeSe in the CB-AFM configuration,
are shown in Fig.~\ref{fig3}.  They are essentially similar to each other and to Fig.~\ref{fig2}(e) for CL-AFM order, 
suggesting that the magnetic ordering of the Fe dopants has negligible effects on the band structure of
the FeSe layers. (There are some noticeable differences; for example, the state nearest the Fermi level around M for the 
FM case is split, reflecting the fact that in the
FM case there is a net magnetic moment and field from the (Li,Fe)OH layer.) 
This insensitivity to the magnetic ordering of the dopants is not
surprising since the Li$_{0.8}$Fe$_{0.2}$OH and FeSe layers interact via weak vdW-like forces.  Our calculations are
consistent with the observation that the electronic state of (Li$_{0.84}$Fe$_{0.16}$OH)Fe$_{0.98}$Se is highly
two-dimensional.
\cite{dong_2015_PRB}

%
\begin{figure}
  \includegraphics[width=0.48\textwidth]{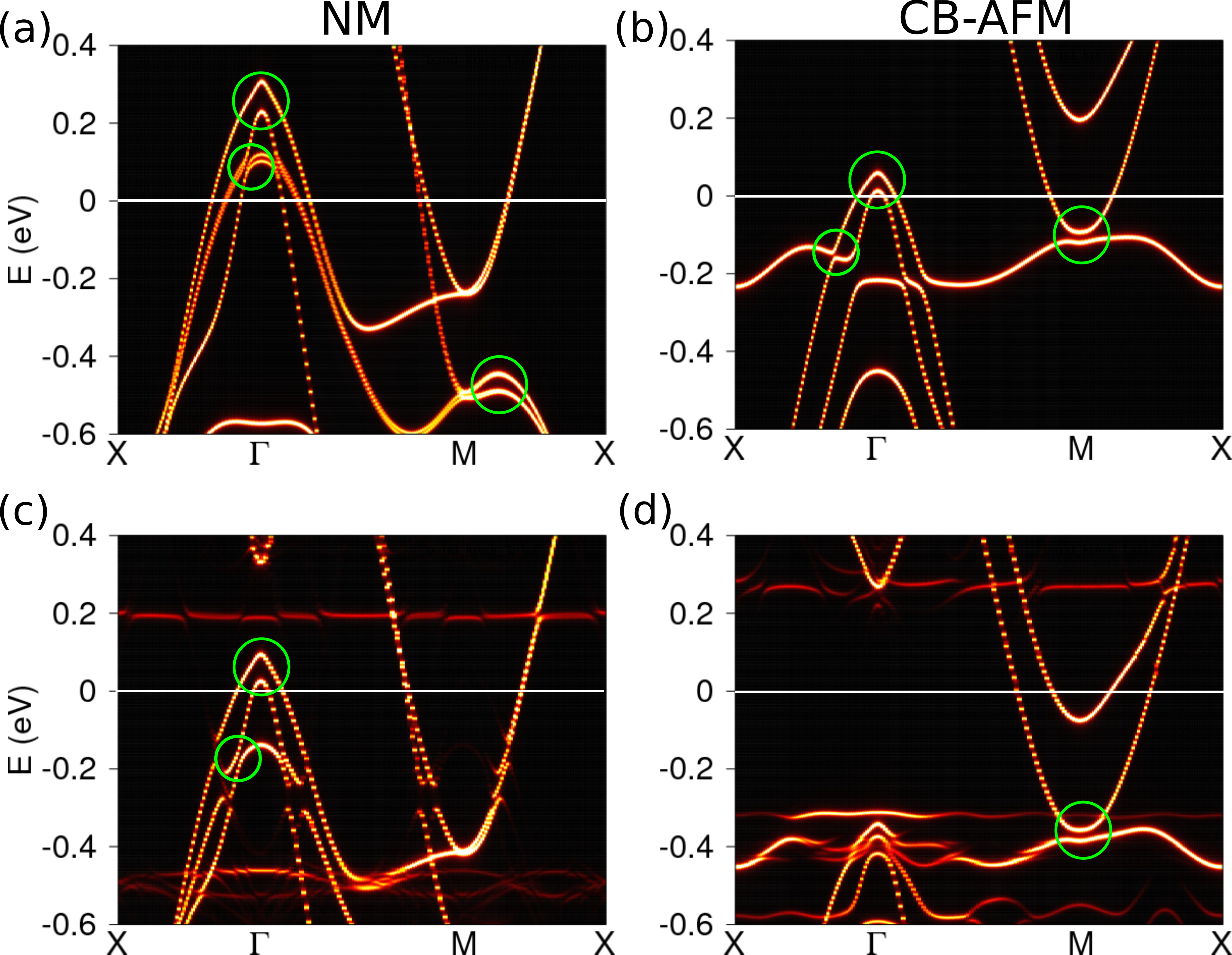}
  \caption{Effect of spin-orbit coupling on the band structure of (a)-(b) (LiOH)FeSe and (c)-(d) (Li$_{0.8}$Fe$_{0.2}$OH)FeSe.
  Pink circles indicate the SOC splittings.
  }
 \label{fig4}
\end{figure}
Spin-orbit coupling is of importance in Fe-based superconductors.
For example, it lifts the degeneracy of the Fe $d_{xz}$ and $d_{yz}$ orbitals at the zone center and 
induces a band splitting of over 50 meV in Fe(Te$_{0.5}$Se$_{0.5}$).\cite{soc-2015}
For FeSe thin films supported on SrTiO$_3$(001), SOC induces a splitting of about 40 meV at the M point.\cite{chen_2016_PRB}
To assess the effect of SOC on (Li$_{1-x}$Fe$_{x}$OH)FeSe,
calculations including SOC were performed for (Li$_{1-x}$Fe$_{x}$OH)FeSe, $x$ = 0 and 0.2, and are shown
in Fig.~\ref{fig4}. 
SOC induces band splittings near $\Gamma$ and M (marked by circles).
For the NM state, a splitting of about 40 meV is obtained for the hole-like bands at $\Gamma$.
For the CB-AFM state, a similar splitting at $\Gamma$ is obtained for (LiOH)FeSe, 
which slightly decreases with the introduction of the Fe dopants (Fig.~\ref{fig4}d); 
at the M point, the splitting is comparable to the one at $\Gamma$ and
is not affected by the Fe dopants.
The effect of SOC on the band structure for the CL-AFM state is less prominent, due in part to the different symmetries of
the wave functions, and are not shown.
Although the SOC-induced splittings are dependent on the specific magnetic configurations -- as are the overall band
structures -- they are not expected to play an important role in the superconductivity of (LiFeOH)FeSe since 
the relevant splitting are well away from the Fermi level.

\subsection{Surface bands of (Li$_{0.8}$Fe$_{0.2}$OH)FeSe(001)}
\begin{figure*}
  \includegraphics[width=0.99\textwidth]{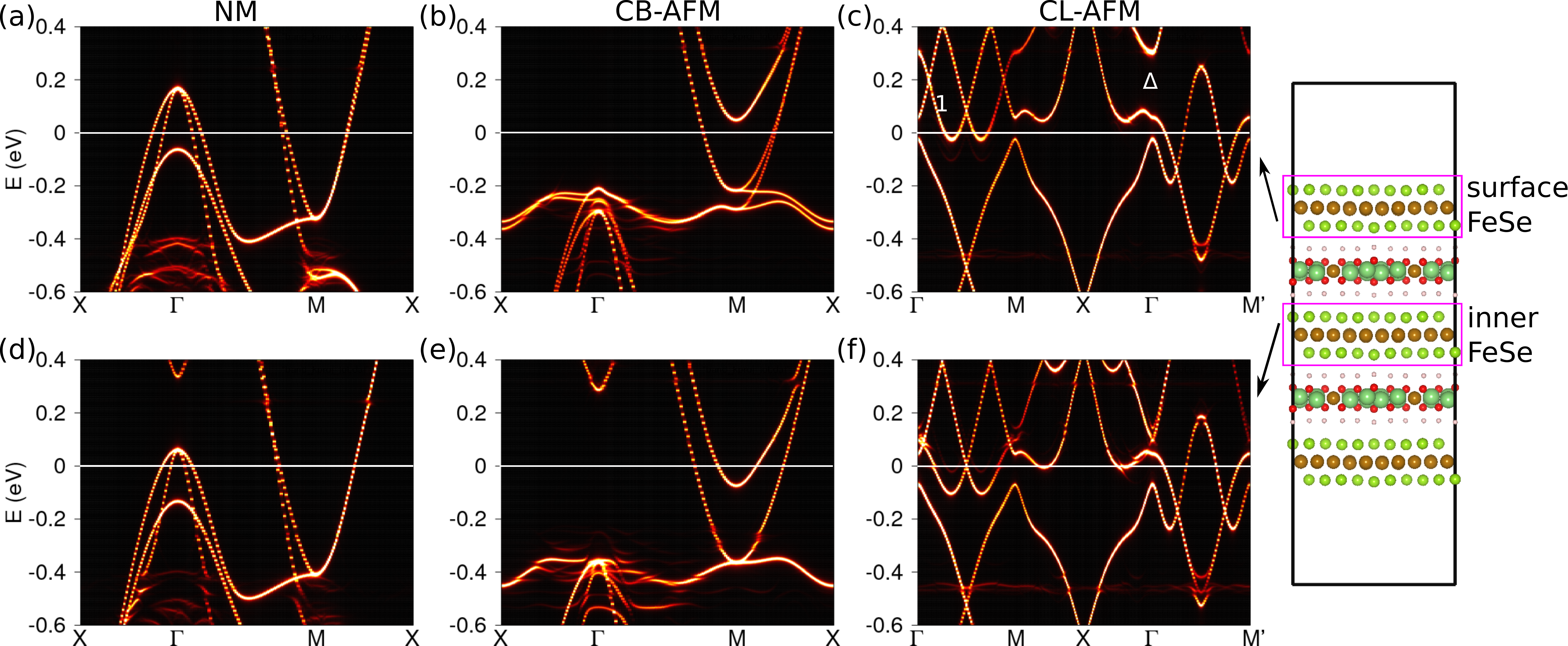}
  \caption{$k$-projected bands for (Li$_{0.8}$Fe$_{0.2}$OH)FeSe(001).
  The Fe dopants are in the CL-AFM configuration and the Fe atoms in the FeSe layers have the NM, CB-AFM and CL-AFM
orderings, respectively. The Fermi level is set to zero.
  }
 \label{fig5}
\end{figure*}

ARPES experiments observe two hole-like bands below E$_\mathrm{F}$ near $\Gamma$ with the top of the band at about $-$50 meV.
\cite{Feng_PRB_2015,zhao_common_2016}
Around the M point, there are parabolic electron-like bands crossing E$_\mathrm{F}$ with the bottom at about $-$60 meV, 
below which there are hole-like bands.\cite{Feng_PRB_2015,zhao_common_2016}
None of the above calculations produce agreement with the experiments,
although the band structure for the CB-AFM state (Fig.~\ref{fig4}(d)) is
similar to the experimental observations in that 
there are no hole pockets around $\Gamma$ and a gap appears at M.
However, there are two electron pockets around M in our calculations instead of the single one observed in the ARPES experiments.
Since ARPES is surface-sensitive (for photons of $\sim$20-22 eV, the expected probing depth is $\sim$5
\AA\cite{arpes-depth}), the experiments of
Refs.~[\onlinecite{Feng_PRB_2015}] and [\onlinecite{zhao_common_2016}] will be sensitive to 
the surface bands of (Li$_{0.8}$Fe$_{0.2}$OH)FeSe, which may differ from those of the bulk.

To investigate possible changes in the band structure of (Li$_{0.8}$Fe$_{0.2}$OH)FeSe at the surface,
we have performed calculations for the (001) surface
modeled by a FeSe-terminated slab (Fig.~\ref{fig5}).
Layer projections were performed to separate the contributions from the surface FeSe and the inner FeSe layer.
Electronic bands were also unfolded using the $k$-projection technique.
While the bands for the inner FeSe layer, Figs.~\ref{fig5}(d-f), are essentially the same as the corresponding 
bulk bands in Figs.~\ref{fig2}(d-f), the surface bands differ, especially in placement relative to the Fermi level. 
In all three cases, the surface bands are shifted up relative to the bulk and closer to the positioning of the bands for
the undoped (LiOH)FeSe case. The simple explanation for this behavior is that the surface FeSe layer is less heavily doped
since it has only one, not two, neighboring Li$_{0.8}$Fe$_{0.2}$OH layers so that the surface has only approximately half
the electron doping compared to the bulk. Note that the applicability of such a simple argument implicitly invokes the
2-D layered nature of both FeSe and LiOH layers, and the weak direct bonding between them. In addition to simple
band filling contributions (Fermi level shifts), there are the band renormalization effects that accompany the band filling, as well as the
standard surface potential shifts arising from ``missing'' neighbors and the modifications resulting from the lowered symmetry at
the surface.

As a consequence of these effects,
the hole pockets around $\Gamma$ for the surface FeSe for the NM configuration are larger, whereas the electron pockets around $M$ shrink.
For the CB-AFM state around M, there is only one electron pocket, with the second one pushed up above the Fermi level. 
In addition, a band splitting of about 70 meV at M (located at about $-$0.3 eV) can be seen in the surface band structure,
related to the broken inversion symmetry of the surface FeSe layer.
The hole-like bands near $\Gamma$ (about $-$0.2 eV) is closer to E$_\mathrm{F}$ for the surface FeSe layer than for the inner FeSe. 
For the CL-AFM configuration, surface effects are also noticeable, arising from similar mechanisms:
Besides the shift of bands relative to E$_\mathrm{F}$, 
the top of band 1 (Fig.~\ref{fig5}(c)) at $\Gamma$ has an upward shift,
giving rise to a larger bandwidth compared to that for the inner FeSe, and
the gap at $\Gamma$ marked by $\Delta$ considerably increased for the surface FeSe layer.

The calculated band structure for the surface FeSe layer in the CB-AFM state (Fig.~\ref{fig5}(b)) has similarities to the
ARPES experiments, in that there is no hole pocket around $\Gamma$, only one electron pocket exists around M, and the size
of the gap between the electron-like bands and the hole-like ones at the M point (along $\Gamma$-M) is comparable to the
experimental value ($\sim$70 meV).  However, there is a discrepancy in the Fermi level between our calculations and the
ARPES results: Our calculations find that E$_\mathrm{F}$ is about 200 meV above the top of the hole-like bands at $\Gamma$, rather
than the 70 meV found in the ARPES experiments.  The discrepancy could be due to defects or smaller Fe doping levels that
would act to reduce the electron doping; the reported (Li$_{0.84}$Fe$_{0.16}$OH)Fe$_{0.98}$Se in
Ref.~[\onlinecite{zhao_common_2016}] would tend to go in that direction.  In addition, direct comparisons of our calculations to the ARPES
experiments is hindered by our neglect of final state effects and strong electron correlations that are believed might play
an important role in the electronic structure of Fe-based superconductors,\cite{PRB.81.014526,yin_2011} particularly
orbital-selective strong renormalization.\cite{orbital_2015} Moreover, SOC was not included in our calculations of the
surface structure for computational reasons since the slab has a large number of atoms (240 atoms).  However, as in the
case of bulk (Li$_{0.8}$Fe$_{0.2}$OH)FeSe, the interesting SOC-induced splittings occur away from the Fermi level, and in
the situations where the lower symmetry at the surface already induces a splitting, the SOC splitting is not
additive,\cite{chen_2016_PRB} and thus the calculated splittings (c.f., Fig.~\ref{fig5}) are not expected to change
dramatically.

\subsection{Effects of Mn and Co dopings}
\begin{figure*}
  \includegraphics[width=0.97\textwidth]{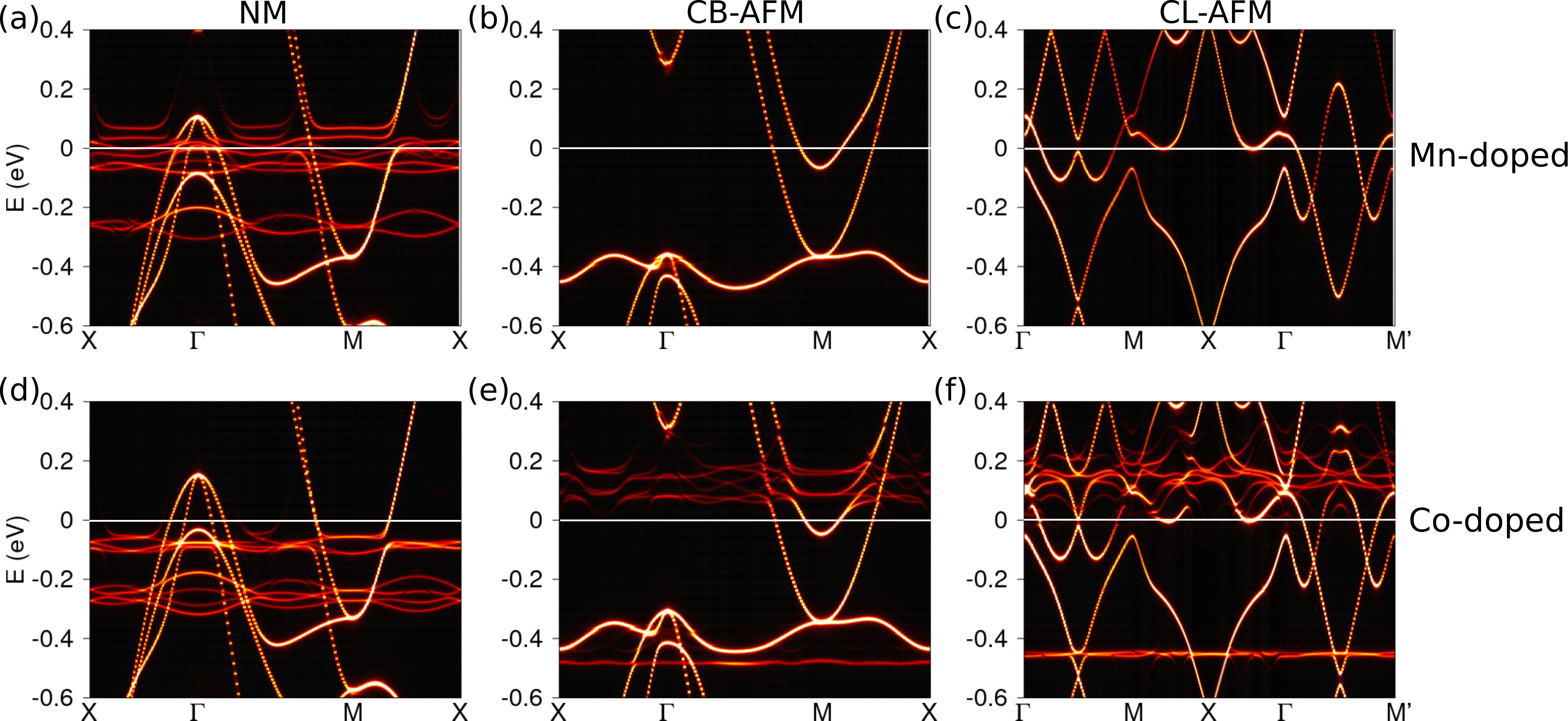}
  \caption{$k$-projected bands for (a)-(c) (Li$_{0.8}$Mn$_{0.2}$OH)FeSe and (d)-(f) (Li$_{0.8}$Co$_{0.2}$OH)FeSe. 
  The dopants in the LiOH layer are in the CL-AFM configurations.
  }
 \label{fig6}
\end{figure*}
\begin{figure}[b]
 \includegraphics[width=0.98\columnwidth]{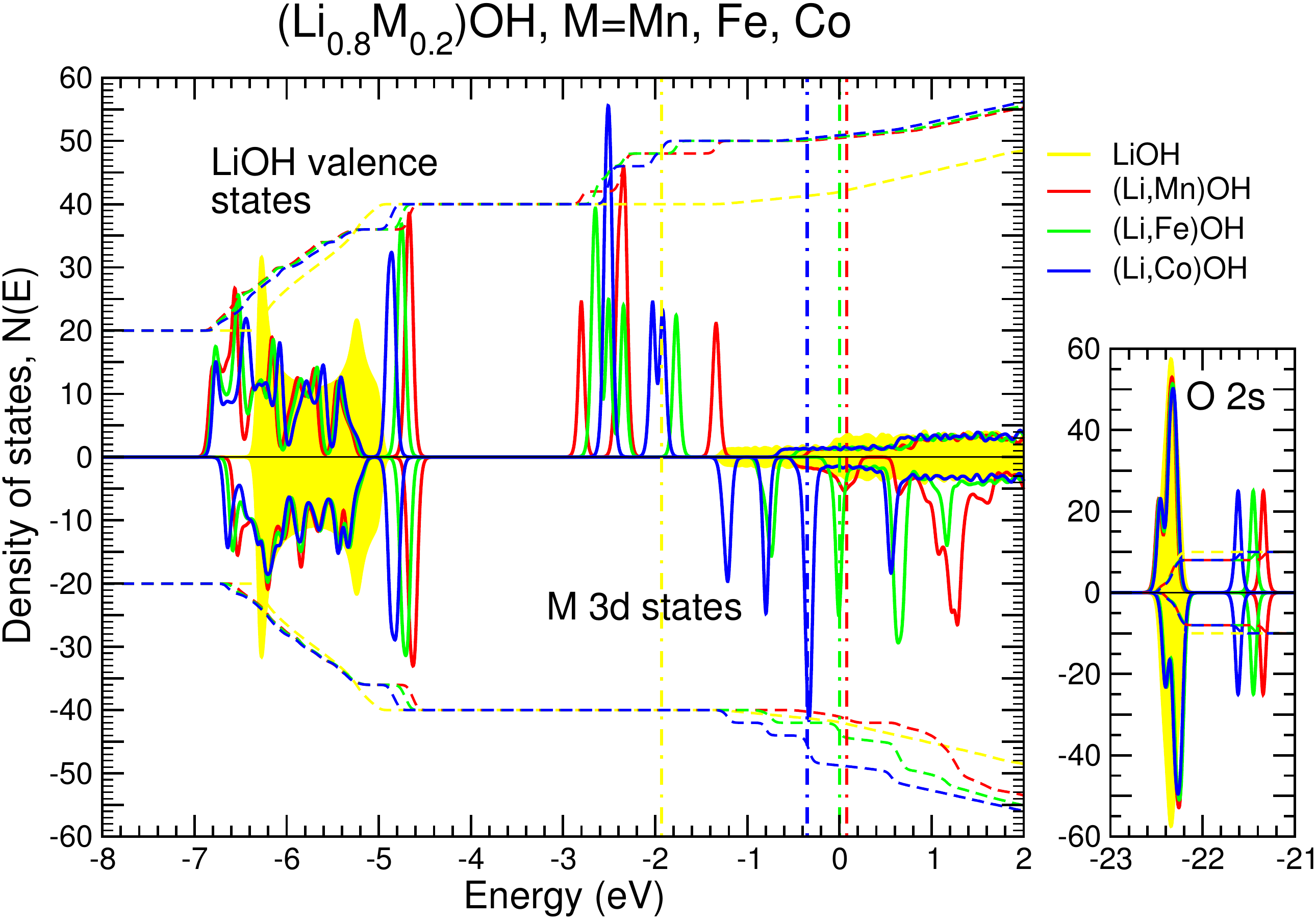}
 \caption{Spin-resolved total density of states (solid lines) for isolated
Li$_{0.8}$M$_{0.2}$OH, M = Li (yellow), Mn (red), Fe (green),
Co (blue), calculated in the $\sqrt{5}\times\sqrt{5}$ supercell for
ferromagnetic ordering of the M moments. (Antiferromagnetic coupling
leads to similar \textit{local} behavior and conclusions regarding the
occupations and position of states, but the FM ordering simplifies the discussion by
separating the spins.) The integrated DOS, N(E), are given by the
corresponding dashed lines. The different calculations were aligned using
the O 2$s$; the calculated Fermi level for each is denoted by the
vertical dashed-dotted lines, with E$\mathrm{F}$ of (Li,Fe)OH setting the overall zero.
}
\label{fig7}
\end{figure}
The critical role of the Fe dopants on the high T$_\mathrm{c}$ superconductivity observed in (Li$_{0.8}$Fe$_{0.2}$OH)FeSe
and the important effects on the band structure of (LiOH)FeSe as revealed by the above calculations raise the question
of how the results depend on choice of dopant, and possibly help in
optimizing the superconducting T$_\mathrm{c}$ in (Li$_{1-x}$M$_{x}$OH)FeSe.
A natural option would be the neighboring elements of Fe, i.e., Mn and Co, 
which by a simple electron counting are expected to give rise to different doping levels to the FeSe layer.
Indeed, previous calculations show that the charge transfer between the LiOH layers and the FeSe layers is enhanced by doping Co,
which may potentially increase the superconducting T$_\mathrm{c}$,\cite{Chenwei_PRB_2015}
although the increased doping is significantly smaller than the change in valence.
To further investigate the doping effect of Mn and Co on the band structure of (LiOH)FeSe, 
we have also performed band unfolding calculations, Fig.~\ref{fig6}, for both (Li$_{0.8}$Mn$_{0.2}$OH)FeSe and (Li$_{0.8}$Co$_{0.2}$OH)FeSe,
using the previously reported relaxed structures\cite{Chenwei_PRB_2015} and with the
dopants in the LiOH layers in the CL-AFM ordering (Fig.~\ref{fig1}(b)).
Overall, the band structures for Mn- and Co-doped (LiOH)FeSe share the features of
(Li$_{0.8}$Fe$_{0.2}$OH)FeSe. The energies of the occupied states at M for CB-AFM are unchanged for the different dopings,
while at $\Gamma$ there are small shifts related to the increased splitting for Co.
The insensitivity of the bands to the dopant can be understood by considering the properties of the isolated LiOH
spacer layer, Fig.~\ref{fig7}, which is an insulator ($\sim$3.6 eV gap). When doped with Mn ($d^5s^2$), Fe ($d^6s^2$), or Co
($d^7s^2$), one of the transition-metal 4$s$ electrons
takes the place of the Li 2$s$, albeit distorting the LiOH bands. The virtual bound states arising from the
local majority $d$ states fall in the gap region and are fully occupied. (These states are hybrids of the 3$d$ orbitals
and the surrounding LiOH states and have finite energy widths, but are isolated in energy. Although there are
small energy differences among the different magnetic configurations, the arguments regarding the filling of minority vs.\
minority states still applies because the $d$-like states are in the gap.) For the minority $d$ orbitals,
the lowest ones filled by the remaining $d$ electrons (0, 1, 2 for Mn, Fe, Co, respectively) are similarly in the
gap. The remaining dopant electron is split between the majority LiOH conduction bands and, in the minority channel, a
virtual bound $d$ state in the LiOH conduction states; it these states that provide the doping to FeSe layer.
Moreover, since the extra electrons going from Mn to Co are accommodated in (minority) gap states, the conduction
states around the Fermi level -- and hence doping
contributions -- are similar for all three dopants. The local minority virtual bound $d$ states are visible in
Figs.~\ref{fig2} and \ref{fig5} for Fe and Co; for Mn, the minority $d$ states are all above the Fermi level and
thus do not show up.
The occupied minority ``gap'' state for Fe and Co doping are seen below the Fermi level. The defect level(s) above
E$_\mathrm{F}$ corresponding to the minority $d$ virtual bound state that is
at the Fermi level for the isolated (Li,M)OH spacer layer; for Co (Fe), this state is doubly (singly) degenerate,
which is reflected in these now unoccupied states.
Thus, since it is the quantity rather than the type of dopant -- Mn, Fe, of Co -- that affects the doping level and
hence electronic structure, it may be possible to tailor the dopants in order to optimize both the FeSe electronic
structure and also the stability\cite{Chenwei_PRB_2015} of system.

\section{SUMMARY}
The band structures of (Li$_{0.8}$M$_{0.2}$OH)FeSe (M = Fe, Mn, Co) have been investigated 
via first-principles supercell calculations. 
The effects of doping 3$d$ transition-metal elements on the band structure of FeSe layers
are similar to those  of interfacial oxygen vacancies in FeSe/SrTiO$_3$.  
The dopants provide electron doping to the FeSe layer, shifting E$_\mathrm{F}$ and
leading to the absence of hole pockets at $\Gamma$ for the checkerboard AFM state.
The doping substantially modifies the profile of the Fe-3$d$ band below the Fermi level, including
renormalizing band widths by a factor of about 0.3.
The calculated bands for the surface layer of CB-AFM FeSe-(Li,M)OH  
are overall consistent with ARPES results, with the caveat that the doping levels and intrinsic defects may
further modify the relative placement of the bands relative to the chemical potential.
However, the band structure of the FeSe layers is insensitive to the magnetic ordering and type of dopants,
suggesting the possibility of tailoring the superconducting properties by modifying the doping.

\begin{acknowledgments}
This work was supported by the U.S. National Science Foundation, Division of Materials Research, DMR-1335215,
the ARO MURI Award No. W911NF-14-0247, 
the National Natural Science Foundation of China (Grants Nos. 11774084, 11504357, 61434002, 11634011), 
and the National Key Basic Research Program of China (Grant No. 2014CB921103).
\end {acknowledgments}

\bibliography{references}
\bibliographystyle{apsrev4-1}
\end{document}